\documentclass[aps,superscriptaddress, preprintnumbers ,twocolumn,showpacs,amsmath,amssymb]{revtex4}
\usepackage{bm}
\usepackage[colorlinks=true,linkcolor=blue]{hyperref}
\usepackage{amsmath}
\usepackage{amssymb}
\usepackage{amsthm}
\usepackage{amsfonts}
\usepackage{times}
\usepackage{enumerate}
\usepackage{latexsym}
\usepackage{ifpdf}
\usepackage{graphicx}
\usepackage{makeidx}
\expandafter\ifx\csname package@font\endcsname\relax\else
 \expandafter\expandafter
 \expandafter\usepackage
 \expandafter\expandafter
 \expandafter{\csname package@font\endcsname}%
 \fi
\usepackage{color}

\begin{document}

\title{Strain-mediated metal-insulator transition in epitaxial  ultra-thin films of NdNiO$_{3}$}
\author{Jian Liu}
%\email{jxl026@uark.edu}
\affiliation{Department of Physics, University of Arkansas, Fayetteville, AR 72701}
\author{M. Kareev}
\affiliation{Department of Physics, University of Arkansas, Fayetteville, AR 72701}
\author{B. Gray}
\affiliation{Department of Physics, University of Arkansas, Fayetteville, AR 72701}
\author{J.W. Kim}
\affiliation{Advanced Photon Source, Argonne National Laboratory, Argonne, IL 60439}
\author{P. Ryan}
\affiliation{Advanced Photon Source, Argonne National Laboratory, Argonne, IL 60439}
\author{B. Dabrowski}
\affiliation{Department of Physics, Northern Illinois University, DeKalb, IL 60115}
\author{J.W. Freeland}
\affiliation{Advanced Photon Source, Argonne National Laboratory, Argonne, IL 60439}
\author{J. Chakhalian}
\affiliation{Department of Physics, University of Arkansas, Fayetteville, AR 72701}
%\author{M. Kareev$^1$, B. Gray$^1$, J.W. Kim$^2$, P. Ryan$^2$, B. Dabrowski$^3$, J. W. Freeland$^2$}
%\author{J. Chakhalian$^1$}

%\affiliation{$^1$Department of Physics, University of Arkansas, Fayetteville, AR 72701}
%\affiliation{$^2$Advanced Photon Source, Argonne National Laboratory, Argonne, IL 60439}
%\affiliation{$^3$Department of Physics, Northern Illinois University, DeKalb, IL 60115}

\begin{abstract}
    We have synthesized epitaxial NdNiO$_{3}$ ultra-thin films in a layer-by-layer growth mode under tensile and compressive strain  on SrTiO$_{3}$ (001)\ and LaAlO$_3$ (001) respectively. A combination of X-ray diffraction, temperature dependent resistivity, and soft X-ray absorption spectroscopy has been applied to elucidate  electronic and structural properties of the samples. In contrast to the bulk NdNiO$_{3}$, the metal-insulator transition under compressive strain is found to be completely quenched,  while the transition remains under the tensile strain albeit modified from  the bulk behavior.
\end{abstract}

%\date{\today}
\maketitle

%\section{Introduction}
%Because of the strong couplings among lattice, orbital and electronic  degrees of freedom, complex oxides exhibit  a rich variety of  electronic and structural phases and, in particular, the Mott metal-insulator transition (MIT). Recent work  focused on  manipulation of strongly correlated electrons in  reduced dimensions along with the intriguing physics associated with surface and interface between two dissimilar oxide materials has opened new prospects for stabilizing unusual quantum phases with the properties not attainable in the bulk \cite{ahn06,Caviglia}. On this route, it is  critical to  understand how confinement, broken symmetry and lattice misfit can alter the MIT  to obtain systems with exquisitely controlled and engineered properties.

Since the Mott metal-insulator transition (MIT) is the hallmark of strongly correlated electron systems, understanding it in the ultra-thin limit is fundamentally and technologically important. The bulk nickelate family, RENiO$_{3}$ (RE = rear earth, e.g. La, Pr, Nd ...), is a prototypical strongly correlated system exhibiting fascinating bandwidth-controlled MIT that is largely attributed to the decrease (increase) in the Ni-O-Ni bond angle with smaller (larger) RE cation \cite{Imada}. In this family, NdNiO$_{3}$ (NNO)\ possesses a  rich set of  properties such as proximity to localized-to-itinerant behavior,   an unusual antiferromagnetic structure, and  a charge-ordered  insulating ground state \cite{Zhou, Garcia1, Scagnoli}.  %Based on the ZAS scheme[refs] NNO is a charge-transfer compound.
With increasing temperature,  NNO undergoes a first order insulator-metal transition and becomes a paramagnetic metal, accompanied by a structural phase transition from monoclinic to orthorhombic \cite{Garcia2}. It is thereby an ideal candidate to investigate the behavior of MIT when subjected to confinement, lattice misfitm and broken symmetry at the interface (surface).

However, due to the low thermodynamic stability of rare earth nickelates, the synthesis of NdNiO$_{3}$ single crystals requires high temperature and high oxygen pressure ($>$100bar) and yields only micron size samples \cite{Lacorre}.  Although, by virtue of the epitaxial stabilization, thin film deposition with relatively low oxygen pressure is now possible, any partial relaxation or 3D island formation during the growth immediately causes oxygen deficiency and secondary phase formation \cite{Gorbenko}. In the absence of the layer-by-layer (LBL)\ growth, the relaxation and the induced non-stoichiometry severely hinders the quantitative interpretation of the effect of epitaxial strain on the electronic structure and is largely responsible for the disagreement among reported results \cite{Catalan0}. For instance, tensile strain has been reported to either suppress\cite{Catalan, Novojilov} or enhance\cite{Eguchi} the MIT  temperature. Similarly controversial results  have been reported for compressive strain \cite{Catalan, Tiwari}. As a result, to obtain fully epitaxial stoichiometric NNO thin films and quantify the role of strain, perfect epitaxy characteristic of  the LBL growth is critical. In addition, LBL growth is   crucial in achieving the high  morphological quality  necessary for hetero-junctions and devices with atomically flat surfaces and interfaces.

In this letter, we report on the  fabrication of NNO  ultra-thin films on SrTiO$_{3}$ (STO) and LaAlO$_3$ (LAO) substrates with a LBL growth mode by pulsed laser deposition. Characterization by atomic force microscopy (AFM), synchrotron based X-ray diffraction (XRD), and resonant X-ray absorption spectroscopy (XAS) reveals full epitaxy and proper stoichiometry.  Temperature dependent transport measurements and XAS  clearly demonstrate that tensile strain preserves the MIT globally and locally while compressive strain suppresses the insulating phase.

%\section{Sample Preparation}
\begin{figure}[t]\vspace{-0pt}
\includegraphics[width=8.5cm]{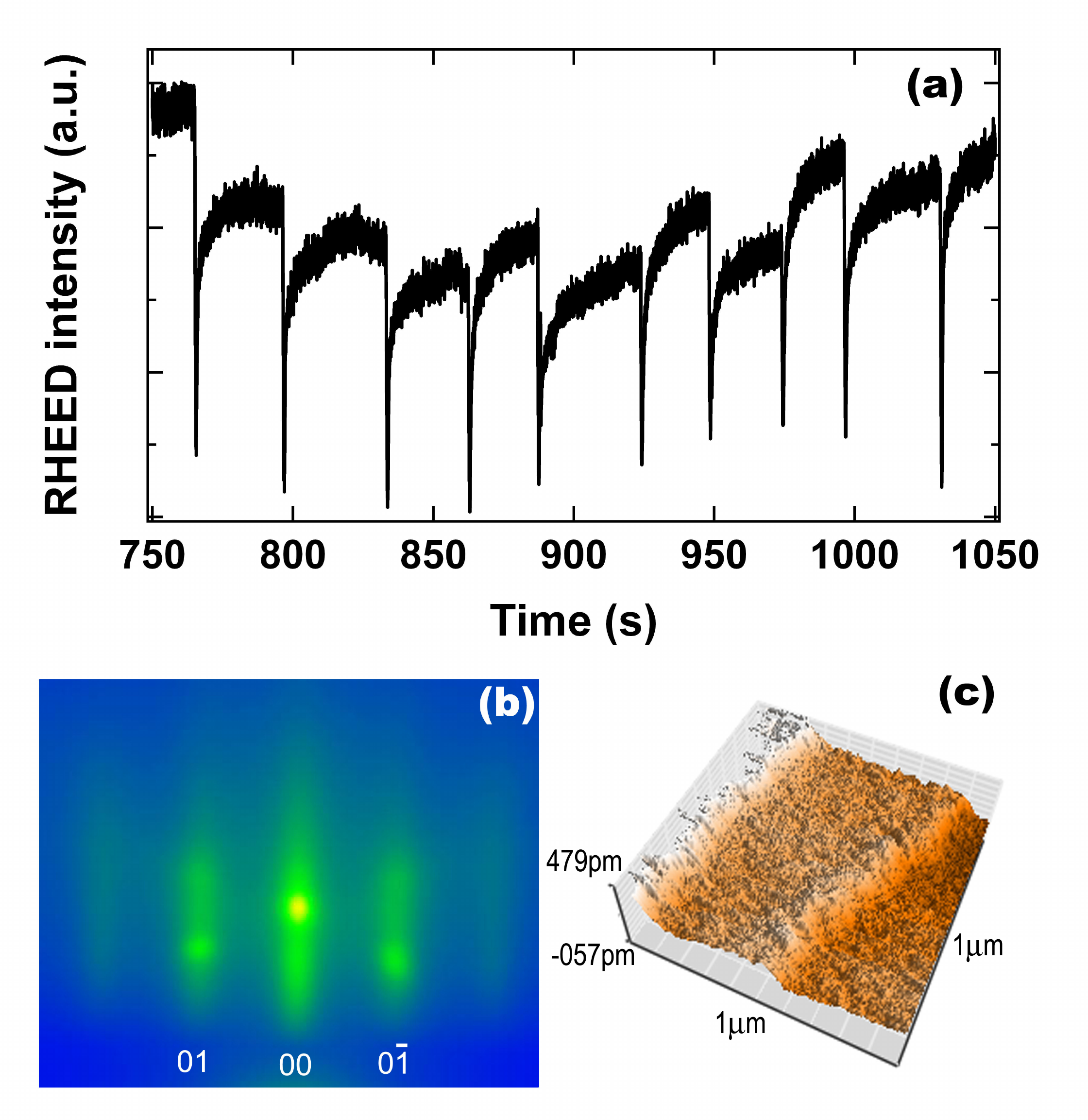}
\caption{\label{RHEED} (a) A representative RHEED specular intensity evolution during the growth of 10uc NNO. (b) RHEED pattern along the [100] direction of pseudo cube. (c) AFM image of a 15uc NNO film grown on STO.}
\end{figure}

High-quality  epitaxial NNO ultra-thin films (10-15uc) were grown on
 STO (001) and LAO (001) single crystal
substrates by pulsed laser deposition with \textit{in-situ}
monitoring by Reflection High Energy Electron Diffraction (RHEED). Due to the lattice mismatch, NNO (3.80\ {\AA}) is subjected to a tensile strain of 2.6\%\ on STO (3.90\ {\AA}), whereas a compressive strain of -0.3\%\ is applied on LAO (3.79\ {\AA}). STO substrates were prepared by our recently
developed chemical wet-etch procedure (`Arkansas treatment') to
attain an atomically flat TiO$_{2}$-terminated surface with a minimal number of surface bound electronic defects\cite{Kareev2}.
The growth temperature was in the range of 670-730$^{  o}C$, while the oxygen partial
pressure was maintained at 75-120 mTorr. After deposition, in order to
maintain the  proper oxygen content, samples were post annealed \textit{in-situ}
 for 30 minutes and cooled down to room temperature in 1 atmosphere of ultra-pure oxygen.

To maintain  the
perfect epitaxy, LBL growth is achieved by
 the interrupted deposition \cite{Blank}, which requires a  rapid laser ablation cycle (up to 30Hz) followed by a prolong time-delay between two successive unit cells. Figure 1(a) shows a representative time-dependent RHEED specular intensity
(RSI), where the full recovery of RSI, characteristic of the perfect LBL growth, can be  seen after each
unit-cell layer. Smooth surface morphology is also evident from the well-defined spots of the ($00$) specular and ($01$) and ($0\bar{1}$) off-specular reflections with streaks in the resulting RHEED pattern (see Fig. 1(b)). AFM imaging (Fig. 1(c)) showed that the sample surface  is  atomically flat with    preserved vicinal steps and typical  surface roughness \textless\ 80pm.

\begin{figure}[t]\vspace{-0pt}
\includegraphics[width=8.5cm]{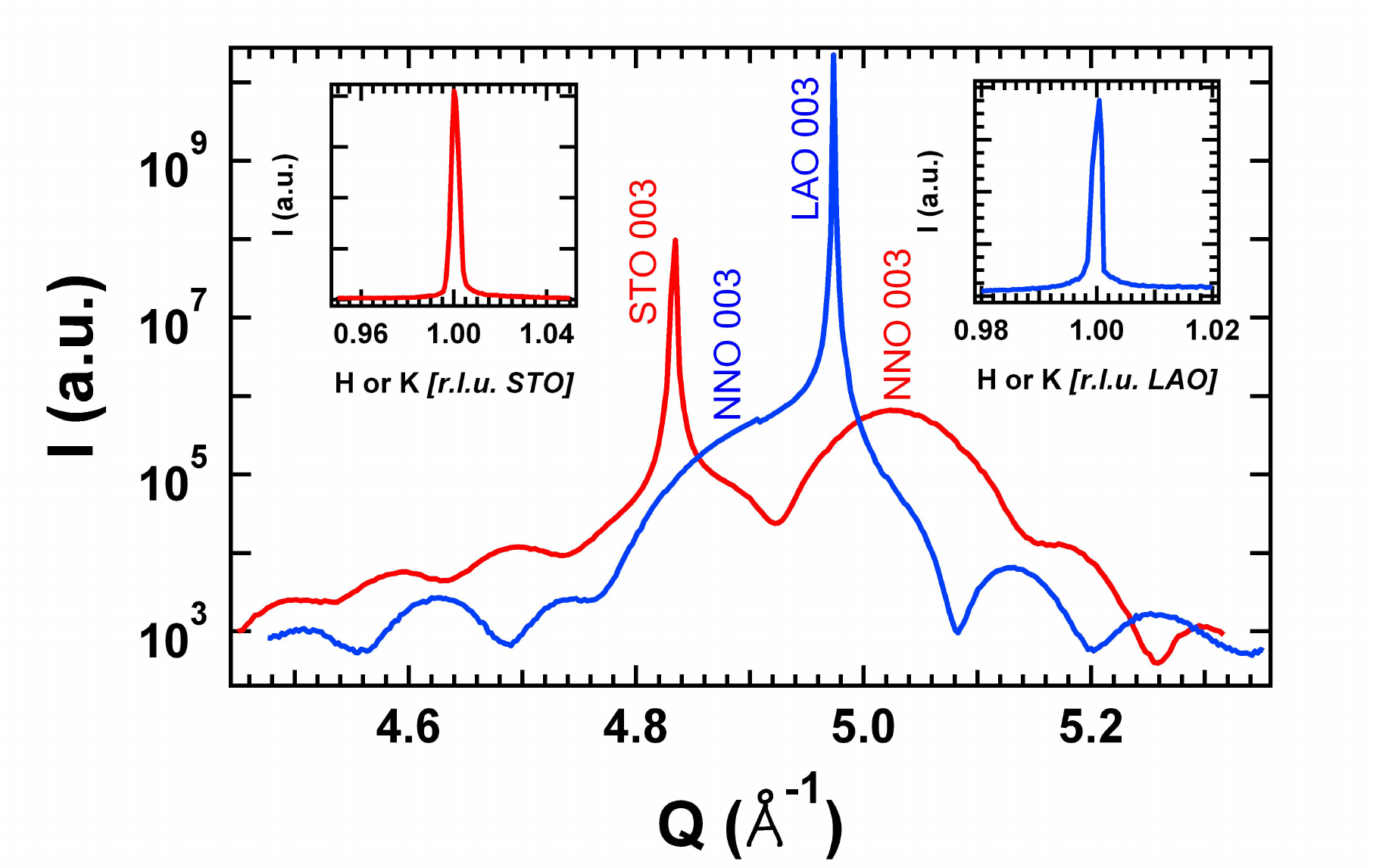}
\caption{\label{XRD} X-ray diffraction scans around the (003) reflections for 15uc
NNO ultra-thin films on STO (red) and LAO (blue). The insets show H=K scans across the NNO (113) reflection of 15uc films on STO (left) and LAO (right).}
\end{figure}

Film structure was determined with synchrotron-based XRD  performed at the 6ID-B beamline
of the Advanced Photon Source (APS). Temperature dependent resistivity was measured using a \textit{van der Pauw} geometry. To probe the local electronic state, detailed spectroscopic measurements were carried out in the soft X-ray
regime in both fluorescence yield (FY) mode and total electron yield
(TEY) mode at the Ni L-edge with 100 meV resolution at the 4ID-C beamline at APS. To obtain precise information on
the Ni valence state, all the spectra were aligned to a NiO
(Ni$^{2+}$) standard which was measured simultaneously with the samples.

Figure 2 shows  the XRD scans around the (003) reflection
for  NNO (15uc) on STO and LAO. Total thickness fringes are clearly seen, indicating coherently grown films.
In addition, their size corresponds to a thicknesses of 15 monolayers, which is in excellent agreement with the number of unit cells deduced from RHEED. The out-of-plane lattice parameters are
3.75{\AA} and 3.84{\AA} on both STO and LAO, respectively. Compared to the bulk, they are consistent
with the expected compression and expansion of the c-axis under bi-axis tensile and compressive strains, respectively. The
insets in Figure 2 shows the H=K scans across the (113) off-specular reflection of the films.
As seen, the in-plane components are perfectly matched to the
substrates, confirming the films  are fully epitaxial.

Temperature dependent resistivity curves are shown in Figure 3. As clearly seen, the sample grown on
STO maintains metallicity  at high temperatures, even though it has only a few monolayers. This is in contrast to the previously  reported results  which show a breakdown of metallicity around few hundred angstroms \cite{Catalan}.  With lowering temperature, a resistivity
minimum is reached at around 180K followed by a rapid increase of resistivity with a thermal hysteresis at lower temperatures. This behavior is characteristic \ of a broadened first order MIT. Notice, the overall increase of resistivity  is over three orders of magnitude and consistent with a bulk-like insulating ground state \cite{Catalan0}. This is opposed to the previous reports, where the resistivity increase under tensile strain was found to be  suppressed to only two orders or less as the film thickness reduces to 10-20nm \cite{Catalan, Kozuka, Goudeau}. On the other hand, compared to the bulk, the observed increase of resistivity within the hysteresis is  more gradual in the ultra-thin structure \cite{Zhou}. In addition to strain, this could also be related to the ultra-thin nature of the film, which has been shown to inhibit and broaden phase transitions of other oxide films\cite{Ziese, Huijben}.  In marked variance to STO, samples grown on LAO are highly metallic within the entire temperature range without any sign of hysteric behavior. This indicates that the bulk MIT is completely quenched by application of the compressive strain.

\begin{figure}[t]\vspace{-0pt}
\includegraphics[width=8.5cm]{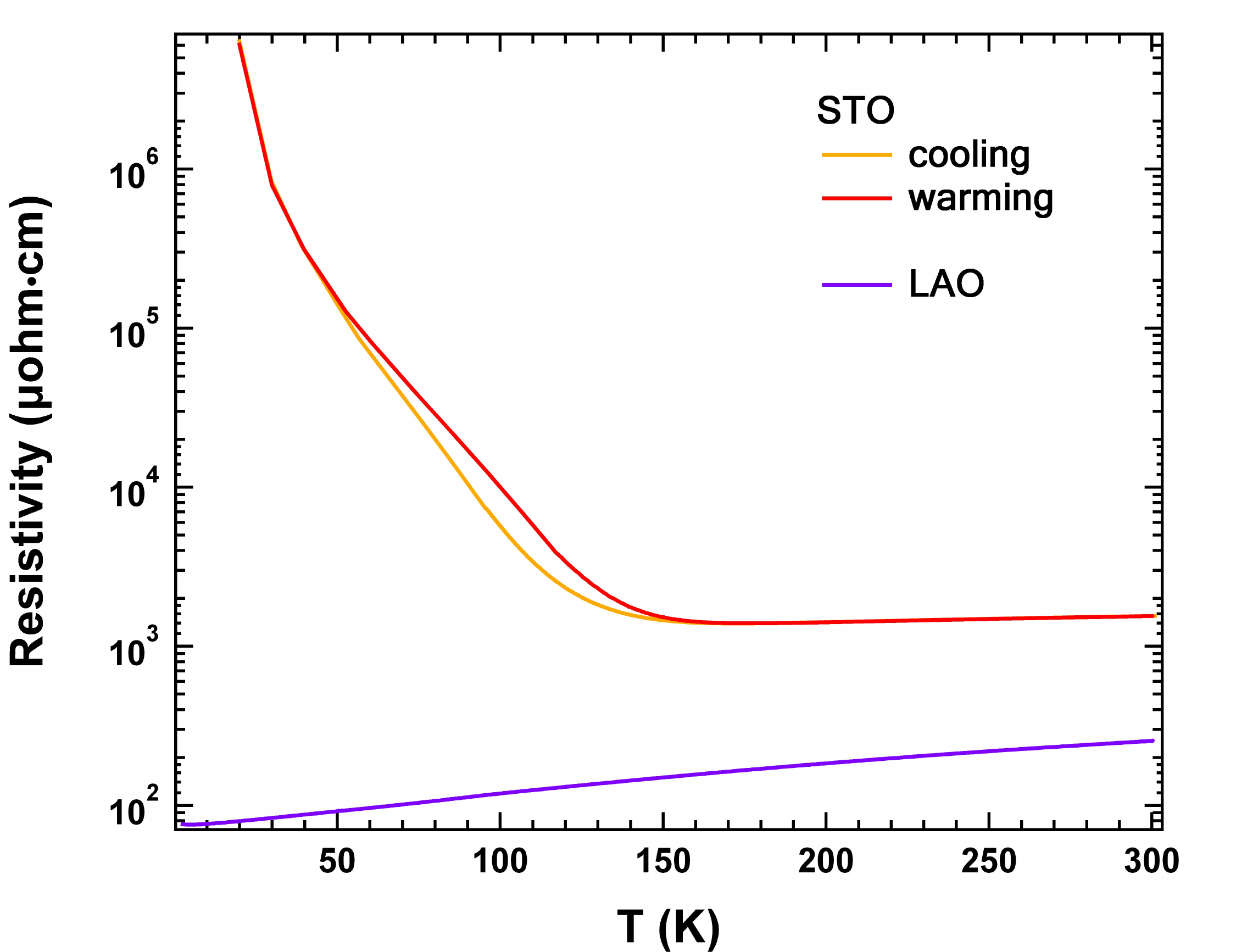}
\caption{\label{transport} Temperature dependent resistivity of 15uc
NNO ultra-thin films on STO and LAO}
\end{figure}

\begin{figure}[t]\vspace{-0pt}
\includegraphics[width=8.5cm]{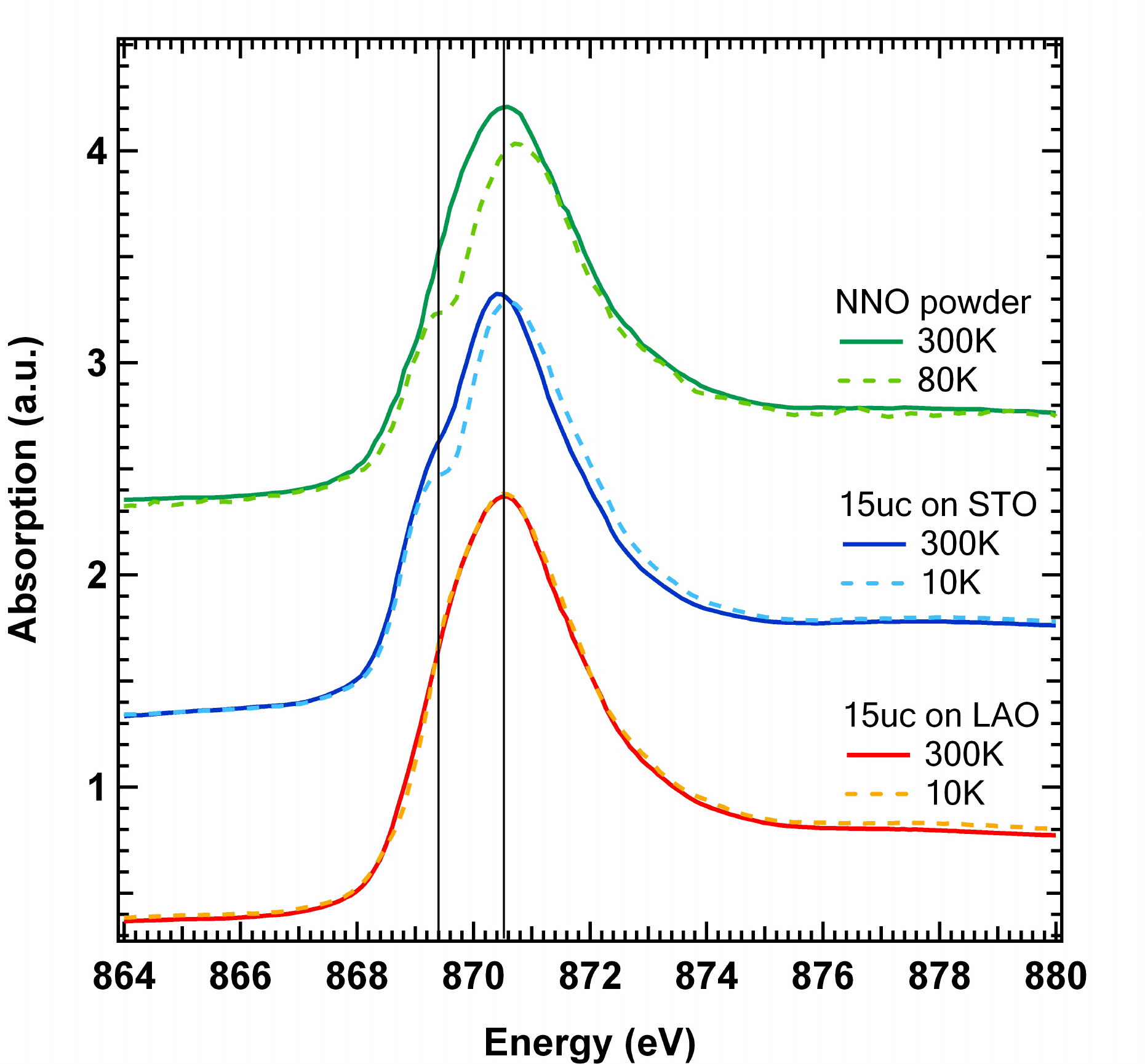}
\caption{\label{XAS} Soft x-ray absorption spectra of 15uc
NNO ultra-thin films on STO and LAO at Ni L$_2$-edge.}
\end{figure}

To elucidate a possible modification of the local Ni electronic structure, XAS spectra of films on STO and LAO were obtained at
300K and 10K, i.e. well below the  bulk MIT temperature of $\sim210$K . Due to the overlapping La M$_4$-edge, the spectra of films on LAO at the
Ni L$_3$-edge is strongly distorted. Instead we focus on the Ni
L$_2$-edge where the distortion is absent. To understand the effects associated with strain and confinement,  we also obtained XAS\ spectra on NNO cold-pressed powder above and below the MIT (300K and 80K) \cite{Zhou}. As seen in Fig. 4, a direct inspection  of the main absorption peak at 870.5eV of both samples at 300K are consistent with  the Ni$^{3+}$ state in the metallic phase\cite{Medarde}. At low temperature, the bulk and film grown on STO show the appearance of a strong multiplet, which is also
observed in other members of   RENiO$_{3}$ series with smaller rare-earth ions in the insulating phase\cite{Piamonteze}. Previous extensive work on the powder sample has revealed  the presence of a charge-order (CO) ground state below the MIT.  In contrast to  this observation, no temperature dependent evolution of the line-shape associated with CO is seen for the sample on LAO which is consistent with the maintained metallic phase even at low temperature.

In the bulk, it has been established that the evolution of MIT is due to the opening of a charge transfer gap with the reduction of the d-band bandwidth \cite{Torrance, Garcia3}. By close analogy, the results above imply, that the suppression of the MIT under compressive strain is likely associated with the closing of the charge transfer gap. It is  interesting to note,  that the MIT can also be suppressed by application of hydrostatic pressure and a complete quenching would require more than 40kBar \cite{Zhou1}. When under tensile strain, the presence of the MIT with a  hysteresis and the bulk-like magnitude of MIT\ strongly  implies  that a sizable charge transfer gap remains in the ground state, despite the existence of electronic modification due to strain and the ultra-thin geometry.

In conclusion, we have developed the LBL\  growth  and synthesized high-quality fully epitaxial  ultra-thin film of NNO on STO and LAO.  Metallicity is maintained on both substrates for films even as thin as 10 unit cells. RHEED, AFM imaging, synchrotron-based XRD and XAS studies have  confirmed the excellent morphology and bulk-like stoichiometry of the samples. Temperature dependent d.c. resistivity and resonant XAS at the Ni L-edge have revealed that the MIT is quenched under the compressive strain of LAO, while the metal-insulator transition  remains under the tensile strain. These findings demonstrate the possibility of  a strain-controlled MIT by ultra-thin film epitaxy for future applications based on heterojunctions of correlated oxides.

The authors acknowledge fruitful discussions with D. Khomskii, A. Millis and G.A. Sawatzky. J.C. was supported by DOD-ARO under the grant
No. 0402-17291 and NSF grant No. DMR-0747808. Work at the Advanced Photon
Source, Argonne is supported by the U.S. Department of Energy, Office of
Science under grant No. DEAC02-06CH11357.

\end{document}